\documentstyle[epsfig]{mn}

\newif\ifAMStwofonts

\ifoldfss
  \ifCUPmtlplainloaded \else
    \NewTextAlphabet{textbfit} {cmbxti10} {}
    \NewTextAlphabet{textbfss} {cmssbx10} {}
    \NewMathAlphabet{mathbfit} {cmbxti10} {} 
    \NewMathAlphabet{mathbfss} {cmssbx10} {} 
  \fi
  \ifAMStwofonts
    \ifCUPmtlplainloaded \else
      \NewSymbolFont{upmath} {eurm10}
      \NewSymbolFont{AMSa} {msam10}
      \NewMathSymbol{\upi}     {0}{upmath}{19}
      \NewMathSymbol{\umu}     {0}{upmath}{16}
      \NewMathSymbol{\upartial}{0}{upmath}{40}
      \NewMathSymbol{\leqslant}{3}{AMSa}{36}
      \NewMathSymbol{\geqslant}{3}{AMSa}{3E}

       \let\le=\leqslant
       \let\ge=\geqslant
    \fi
  \fi
\fi 

\ifnfssone
  \newmathalphabet{\mathit}
  \addtoversion{normal}{\mathit}{cmr}{m}{it}
  \addtoversion{bold}{\mathit}{cmr}{bx}{it}
  \newmathalphabet{\mathbfit} 
  \addtoversion{normal}{\mathbfit}{cmr}{bx}{it}
  \addtoversion{bold}{\mathbfit}{cmr}{bx}{it}
  \newmathalphabet{\mathbfss} 
  \addtoversion{normal}{\mathbfss}{cmss}{bx}{n}
  \addtoversion{bold}{\mathbfss}{cmss}{bx}{n}
  \ifAMStwofonts
    \ifCUPmtlplainloaded \else
and
your
      \UseAMStwoboldmath
      \makeatletter
      \new@mathgroup\upmath@group
      \define@mathgroup\mv@normal\upmath@group{eur}{m}{n}
      \define@mathgroup\mv@bold\upmath@group{eur}{b}{n}
      \edef\UPM{\hexnumber\upmath@group}
      \new@mathgroup\amsa@group
      \define@mathgroup\mv@normal\amsa@group{msa}{m}{n}
      \define@mathgroup\mv@bold\amsa@group{msa}{m}{n}
      \edef\AMSa{\hexnumber\amsa@group}
      \makeatother
      \mathchardef\upi="0\UPM19
      \mathchardef\umu="0\UPM16
      \mathchardef\upartial="0\UPM40
      \mathchardef\leqslant="3\AMSa36
      \mathchardef\geqslant="3\AMSa3E

       \let\le=\leqslant
       \let\ge=\geqslant
    \fi
  \fi
\fi 

\ifnfsstwo
  \DeclareMathAlphabet{\mathbfit}{OT1}{cmr}{bx}{it}
  \SetMathAlphabet\mathbfit{bold}{OT1}{cmr}{bx}{it}
  \DeclareMathAlphabet{\mathbfss}{OT1}{cmss}{bx}{n}
  \SetMathAlphabet\mathbfss{bold}{OT1}{cmss}{bx}{n}
  \ifAMStwofonts
    \ifCUPmtlplainloaded \else
      \DeclareSymbolFont{UPM}{U}{eur}{m}{n}
      \SetSymbolFont{UPM}{bold}{U}{eur}{b}{n}
      \DeclareSymbolFont{AMSa}{U}{msa}{m}{n}
      \DeclareMathSymbol{\upi}{0}{UPM}{"19}
      \DeclareMathSymbol{\umu}{0}{UPM}{"16}
      \DeclareMathSymbol{\upartial}{0}{UPM}{"40}
      \DeclareMathSymbol{\leqslant}{3}{AMSa}{"36}
      \DeclareMathSymbol{\geqslant}{3}{AMSa}{"3E}

       \let\le=\leqslant
       \let\ge=\geqslant
    \fi
  \fi
\fi 

\ifCUPmtlplainloaded \else
  \ifAMStwofonts \else 
    \def\upi{\pi}
    \def\umu{\mu}
    \def\upartial{\partial}
  \fi
\fi

\title{Early chemical enrichment of the universe and the role of very massive pop III stars}
\author[F. Matteucci, F. Calura]
       {F. Matteucci$^{1}$,  
        F. Calura$^{1}$\thanks{E-mail: fcalura@ts.astro.it} \\
        (1) Dipartimento di Astronomia-Universit\'a di Trieste, Via G.
B. Tiepolo
	11, 34131 Trieste, Italy\\
	 }
	
\date{Accepted for publication.........}

\pagerange{\pageref{firstpage}--\pageref{lastpage}}
\pubyear{2004}

\begin{document}

\maketitle

\label{firstpage}
\begin{abstract}
In this paper the role of very massive pop III stars in the chemical 
enrichment of the early universe is discussed.  
We first compare our predictions with
the abundance ratios measured in the high redshift 
Lyman- $\alpha$ forest to check  whether they are 
compatible with the values predicted by assuming that the early universe 
was enriched by massive pop III stars. We conclude that to explain the 
observed C/Si ratio in the intergalactic medium, a contribution 
from pop II stars to carbon enrichment is necessary, already at redshift z=5. 
We then evaluate the number of Pair-Instability Supernovae 
($SN_{\gamma \gamma}$) required to enrich the universe to the 
critical metallicity $Z_{cr}$, i.e. the metallicity value which causes 
the transition from a very massive star regime 
($m > 100 M_{\odot}$) 
to a lower mass regime, similar to the one characteristic of the 
present time ($m < 100 M_{\odot}$). It is found that between 110
and 115 $SN_{\gamma \gamma}$
are sufficient to chemically enrich a cubic megaparsec of the intergalactic 
medium at high redshift for a variety of initial mass functions.
The number of ionizing photons provided by these $SN_{\gamma \gamma}$ and
also by the pop III stars ending as black holes 
was computed and we conclude that there are not enough photons to reionize 
the universe, being down by at least a factor of $\sim$ 3. 
Finally, we calculate the abundance ratios generated by pop III stars 
and compare it with the ones observed in low metallicity 
Damped Lyman-$\alpha$ systems (DLAs). We suggest 
that pop III stars alone 
cannot be responsible for the abundance ratios in these objects and that 
intermediate 
mass pop II stars must have played an important role especially
in enriching DLAs
in nitrogen.

\end{abstract} 


\section{Introduction}
The large-angle polarization 
anisotropy of the cosmic microwave background (CMB), recently observed 
by the WMAP experiment, can  be used to constrain 
the total production of ionizing photons from the first stars 
(e.g. Cen 2003 a,b; Ciardi et al. 2003), thus indicating  a 
substantial early activity of massive star formation at redshifts $z \ge 15$.
It has  also been suggested  
(e.g. Songaila, 2001, Schaye et al. 2003) that massive 
pop III stars (namely the very first stars with no metals)
could have been responsible for the production of C and Si 
abundances measured in the intergalactic medium (IGM) at high redshift. 

The possible existence of very massive pop III stars, 
is suggested by
recent studies of the collapse of the first cosmic structures which 
have demonstrated that 
a quite large mass range for the proto-galactic clumps seems plausible, 
$10^{2}-10^{4} M_{\odot}$ (Schneider et al. 2002 and references therein). 
Since the physical conditions in the early universe seem to prevent 
the further 
collapse of these heavy clumps, hence the conclusion is that 
the first stellar objects had masses much larger than those of  
present-day stars ($M \ge 100 M_{\odot}$, Bromm \& Larson 2004 and references 
therein). 

As reported by Ferrara \& Salvaterra (2004), these very massive stars should 
continue to form until a critical metallicity is reached by the gas.  
At this critical value, the star formation mode would experience a 
transition from the formation of high mass objects to the
formation of low mass stars, which we observe today. The metallicity 
influences the star formation mode in the following way: for a 
metal-free gas, the only 
efficient coolant is molecular hydrogen. Clouds with a mean metallicity 
$Z=10^{-6} Z_{\odot}$ follow the same evolution in density and temperature 
(the main parameters influencing the Jeans mass)
of the gas with primordial chemical composition. For metallicities larger
than $10^{-4}Z_{\odot}$, the fragmentation 
proceeds further until the density  is $\sim  10^{13} cm^{-3}$ and the 
corresponding Jeans mass is of the order of $10^{-2} M_{\odot}$.
Schneider et al. (2002) concluded that the critical metallicity $Z_{cr}$ 
lies in the range $(10^{-6}-10^{-4}) Z_{\odot}$.

The evolution and 
nucleosynthesis of zero-metal massive and very massive stars has been 
computed since the early eighties 
(e.g. Carr et al. 1982; Ober et al. 1983; 
El-Eid et al. 1983) and has continued until recently 
(e.g. Woosley \& Weaver, 1995; Heger \& Woosley 2002; Umeda \& Nomoto, 2002;
Heger et al. 2003; Chieffi \& Limongi 2004).

In this paper, we test whether very massive pop III stars can 
explain the high redshift abundances observed in the 
IGM. 
According to the previous discussion,  we assume that the very first objects 
polluting 
the universe had masses $m> 100 M_{\odot}$. 
In particular, we compare the results obtained by adopting the
most recent yield calculations for pop III stars, such as  the yields 
by Heger \& Woosley (2002, HW02) and Umeda \& Nomoto (2002, UN02), with the 
observations of the high redshift IGM.
We compute the number of ionizing $SN_{\gamma \gamma}$
necessary to pollute the IGM as well as the number of 
ionizing photons per baryon that pop III stars can produce. 

Then, we study how pop III nucleosynthesis could have affected the Damped 
Lyman-$\alpha$ systems (DLAs).
To do that, we take into account a scenario of pop III stars containing
also  masses 
in the range $12 \le m/M_{\odot} \le 100$. 
This assumption is motivated by the fact that DLAs do not have metallicities 
below $Z= 10^{-2.5} Z_{\odot}$, hence it is plausible that 
they have been mainly enriched by stars with 
masses $m \le 100 M_{\odot}$ which formed at metallicities larger than 
$Z_{cr}$.    

The paper is organized as follows: in section 2 we describe the chemical
evolution modelling and the adopted stellar yields, in section 3 we describe 
our results and compare them with observations. Finally, in section 4 some 
conclusions are drawn.

\section{Basic ingredients for chemical evolution}
\subsection{The initial stellar mass function}

For the stellar initial mass function (IMF), we adopt a single power 
law according to the formula: 
\begin{equation}
\phi(m) = \phi_{0} m^{-(1+x)} 
\end{equation} 

The IMF is normalized as:

\begin{equation}
\int^{M_{up}}_{M_{low}}{m \, \phi(m) \, dm}=1
\end{equation}

where $M_{up}$ and $M_{low}$ represent the upper and lower mass limit, 
respectively. 
We adopt two mass ranges for pop III stars: $10^{2}-10^{3} M_{\odot}$ 
and 12-270 $M_{\odot}$. This latter one is considered 
in order to calculate the abundance 
ratios in DLAs. 
We explore three values for the index $x$, i.e. $x=1.35$, $x=0.95$ and, 
to test the hypothesis of an extremely flat IMF, $x=0.5$.

\subsection{Stellar nucleosynthesis}

The two main sets of stellar yields used here are from HW02 and UN02.

In particular, HW02 provided yields for zero metallicity stars in the 
range 140-260 $M_{\odot}$, whereas UN02 for zero-metallicity stars in the 
mass ranges 13-30$M_{\odot}$ and 150-270$M_{\odot}$. 
For masses between 40 and 140 $M_{\odot}$ and larger than 270$M_{\odot}$,
the objects collapse entirely into 
black holes without any ejection of matter.
Therefore, according to the prescriptions by HW02 and UN02, 
we assume that the only objects 
contributing to the chemical pollution of the early universe had masses 
in the ranges 
$ 12 \le m/M_{\odot} \le 40$ and $140 < m/M_{\odot} \le  270$.
When adopting HW02 yields and a mass range 12-260$M_{\odot}$ we use the yields for stars of zero-metallicity in the mass range 12-40$M_{\odot}$ 
by Woosley \& Weaver (1995). 
For the solar abundances, we adopt those of Anders \& Grevesse (1989).

\subsection{Chemical evolution modelling}

The question to ask is: how many 
hypothetical pop III supernovae are necessary to bring the IGM to 
the level of the
critical metallicity?
The values for the critical metallicity necessary 
to enter a regime of normal star formation are:
$Z_{cr} = (10^{-6} -10^{-4})Z_{\odot}$ 
(Bromm \& Larson 2004; Ferrara \& Salvaterra 2004). 

In order to compute the number of $SN_{\gamma \gamma}$ required to enrich 
the universe at the critical metallicity, which we assume to be 
$Z_{cr}=10^{-4}Z_{\odot}$,
we use a simple analytical approach.  

We assume that the metals are ejected from the first stars directly into the 
IGM, homogeneously polluting it.  
According to the approximation of instantaneous recycling (IRA), the 
metallicity $Z$ in a closed-box system (in this case 
the high-redshift universe) 
is given by: 
\begin{equation}
Z = y_{Z} \cdot ln (M_{tot} / M_{gas})  \label{mu} 
\end{equation} 
where $M_{tot}=M_{*} +M_{gas}$, with $M_{*}$ being the mass which has been 
locked up into stars,
$M_{gas}$ being the mass of gas and $y_{Z}$ the yield of metals
per stellar generation, as defined by 
Tinsley (1980), and expressed as:

\begin{equation}
y_{Z} = \frac{1}{1-R} \int_{m_{TO}}^{M_{up}} m \, p_{Zm} \, \phi(m) dm  \label{yield} 
\end{equation}

which depends on the stellar IMF, $\phi(m)$, and on the stellar yield 
$p_{Zm}$, i.e. the fraction of the stellar mass ejected 
in the form of all the newly created heavy elements by a star of 
initial mass $m$. 
Generally, the mass $m_{TO}=1 M_{\odot}$ is the globular cluster 
``turnoff'' mass 
for a normal IMF, but in our case is either $100 M_{\odot}$ 
or $12 M_{\odot}$. 

The quantity $R$ is the returned fraction, defined as: 
\begin{equation}
R  = \int_{m_{TO}}^{M_{up}} (m-m_{rem}) \, \phi(m) dm
\end{equation}

with $m_{rem}$ being the mass of the remnant of a star of initial mass $m$. 
For stars more massive 
than 
$100 M_{\odot}$, 
we assume that $m_{rem}=0$ for objects with masses 
$140 \le m/M_{\odot} \le 270$ (complete explosion), and  $m_{rem}=m$ 
elsewhere.

From equation (3), we can easily obtain the gas mass fraction 
$\mu=M_{gas}/M_{tot}$ as a function of $y_{Z}$ and $Z$: 

\begin{equation}
\mu = exp (- Z / y_{Z} ) 
\end{equation}

\section{Results}
\subsection{The number of $SN_{\gamma \gamma}$}

In table 1  we show the returned fraction R and the yields per stellar 
generation obtained for the different IMFs. 
In column 1 we show the IMF index $x$, in column 2 the returned fraction and 
in columns 3 and 4 the yields per stellar generation 
calculated by adopting HW02 and UN02 yields, 
respectively. 
In column 5 we report the arithmetic average of the two $y_{Z}$, 
which we adopt, given the similarity of the two sets of yields, to 
calculate the ratio $\mu=M_{gas}/M_{tot}$ (through eq. (6)). 
We define the quantity $\mu_{*}=1-\mu$ as the fraction of matter which 
has been locked up into stars: in particular, in this case it represents only the mass
locked up in massive black holes. The fraction of mass ejected from stars into the IGM be $\mu_{ej}$. 

\begin{center}
\begin{table*}
\caption[]{The yields of metals per stellar generation obtained for the 
three different IMFs and the stellar mass range $100-1000 M_{\odot}$. 
In colum 1 we show the IMF index $x$, in column 2 the returned fraction,
in columns 3 and 4 the total yields per stellar generation calculated adopting HW02 and UN02 yields, respectively, 
and in column 5 the arithmetic average between the two. }
\begin{tabular}{l|l|l|l|l}
\noalign{\smallskip}
\hline
\hline
\noalign{\smallskip}
IMF & $R$ & $y_Z$   (HW02) &   $y_Z$    (UN02) &  $<y_Z>$\\ 
\noalign{\smallskip}
\hline
\noalign{\smallskip}
x=1.35  &   0.228  & 0.13     &    0.16     &   0.145   \\
x=0.95  &   0.131  & 0.07     &    0.08     &   0.080    \\
x=0.50   &  0.050  & 0.02     &    0.03     &   0.030    \\
\hline					
\hline					 
\end{tabular}				
\end{table*}
\end{center}
Then the total mass fraction which has been processed into stars is 
defined as: 
\begin{equation}
\mu_{**}= \mu_{ej} + \mu_{*}
\end{equation}
Therefore, we can write:
\begin{equation}
{\mu_{*} \over \mu_{**}}= 1 - R 
\end{equation}
If we now substitute into this equation the expression of $\mu_{**}$ 
from  eq. (7), we obtain:
\begin{equation}
\mu_{ej}=\mu_{*} ({R \over 1- R})
\end{equation}

Since all the $\mu$ quantities are normalized to $M_{tot}$, we define 
the total mass.
As the total mass, $M_{tot}$, we consider the baryon density $\rho_{b}$
by taking into account a unitary volume of the universe of 1$Mpc^{3}$. 
By 
adopting  a baryon density in units of the critical density  
$\Omega_{b}=0.02 h^{-2}$, as suggested by D measurements is QSO absorbers 
(O'Meara et al. 2001) and by WMAP (Spergel et al. 2003), we obtain 
$\rho_{b}=5.55 \times 10^{9} M_{\odot} Mpc^{-3}$. 

We further assume that, at the time when it is reached, the critical 
metallicity $Z_{cr}$ represents the average metallicity 
of the IGM. 

The density of matter exploded as $SN_{\gamma \gamma}$, expressed as a 
function of the critical metallicity $Z_{cr}$ 
is hence given by first substituting $\mu$ in eq. (6) with $1- \mu_{*}$
and then $\mu_{*}$ with the right side of eq. (9), giving
$\mu_{ej}$ which represents the fraction of mass which 
exploded as $SN_{\gamma \gamma}$. 

Finally, by substituting $M_{tot}$ with $\rho_{b}$ and 
$M_{SN_{\gamma \gamma}}$ with 
$\rho_{\gamma \gamma}$ we obtain: 
\begin{equation}
\rho_{\gamma \gamma}= [1-exp(-Z_{cr}/y_{Z} )] \cdot {R \over 1-R} \cdot \rho_{b} 
\end{equation}
If we consider that a $SN_{\gamma \gamma}$ has a typical mass of 
$200 M_{\odot}$, it is easy to retrieve the number 
$N_{\gamma \gamma}$ of $SN_{\gamma \gamma}$  necessary to enrich a volume 
of 1 $Mpc^{3}$ to the critical metallicity $Z_{cr}$: 
\begin{equation}
N_{\gamma \gamma} = \rho_{\gamma \gamma}/200 M_{\odot}
\end{equation}

In figure 1 we present a plot of $N_{\gamma \gamma}$ as a function of the 
critical metallicity $Z_{cr}$.

\begin{figure*}
\centering
\vspace{0.001cm}
\epsfig{file=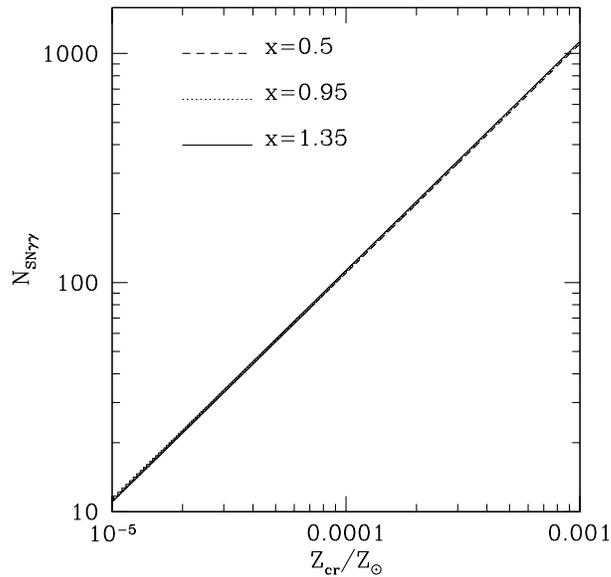,height=8cm,width=8cm}
\caption[]{Predicted number $N_{\gamma \gamma}$ of  $SN_{\gamma \gamma}$ necessary to enrich the IGM at the critical 
metallicity $Z_{cr}$ as a function of $Z_{cr}/Z_{\odot}$.  
\emph{Solid line}: results obtained with an IMF index $x=1.35$; 
\emph{dotted line}: results obtained with an IMF index $x=0.95$;  
\emph{dashed line}: results obtained with an IMF index $x=0.5$. 
}	
\end{figure*}

In table 2 we show the comoving density of matter  which exploded as 
$SN_{\gamma \gamma}$ and the  
number of $SN_{\gamma \gamma}$ required to enrich the IGM at the critical 
metallicity $Z_{cr} = 10^{-4} Z_{\odot}$.

\subsection{Ionizing photons per baryon}

At this point, we can compute the number of ionizing photons per baryon 
produced by both the $SN_{\gamma \gamma}$ necessary to enrich the IGM
and  the progenitors of those very massive pop III stars 
which do not enrich the 
IGM since they collapse directly into black holes.

We rely on the results of
Schaerer (2002) who has calculated the number of ioizing photons as a 
function of the initial mass $N_{ion,*}(m)$  
of the pop III objects, in the range  
$80-1000 M_{\odot}$. 
In our case, we consider only objects in the mass range $100-1000 M_{\odot}$. 

We compute the average number of ionizing photons produced by a typical 
pop III star,
according to:\\ 
\begin{equation}
<N_{ion}> =\frac{\int_{100}^{1000} N_{ion,*}(m) \, \phi(m) dm}{\int_{100}^{1000} \phi(m) dm}
\end{equation}
and show our results, for the 3 different IMFs used here, in 
column 2 of table 3.

The total number of ionizing photons per $Mpc^{-3}$ is then given by: 
\begin{equation}
N_{ion}= <N_{ion}>   \cdot  N_{popIII} 
\end{equation} 
where $N_{popIII}$ is the total number of pop III objects formed 
per $Mpc^{3}$, including both $SN_{\gamma \gamma}$ and black holes. 
To calculate this number, we normalize the IMF in number between 
100 and 270 $M_{\odot}$ by means of the already known
number of $SN_{\gamma \gamma}$ exploded per $Mpc^{3}$, 
which enriched the universe to the critical metallicity 
$Z_{cr}=10^{-4} Z_{\odot}$. 
As a consequence, by integrating the IMF in number with this  normalization in 
the range 100-1000 $M_{\odot}$, 
we obtain the total number of pop III objects born per $Mpc^{3}$, i.e. 
284, 282, 303 for $x=1.35$, $x=0.95$, $x=0.5$, respectively. 

In table 3, column 3, we show the total number of ionizing photons 
(per $Mpc^{3}$) for the three different IMFs considered here. 
At this point, to calculate the number of ionizing photons per baryon, 
we need the value for the baryon density discussed before. 
 In particular, we divide $\rho_{b}$ by the proton mass in solar masses ($m_p=
8.4 \cdot 10^{-58} M_{\odot}$) and obtain the total number of baryons per $Mpc^{3}$:
\begin{equation}
n_{b}=6.6  \cdot 10^{66}
\end{equation} 
The predicted numbers of ionizing photons per baryon are given in column 
4 of table 3. As one can see, at best we predict 1.6 photons per baryon, 
below the estimated number of 5-20 (Sokasian et al., 2003) necessary to 
reionize the universe. 
This means that pop III objects play a non-negligible role in the IGM 
reionization, 
although the contribution of other photon sources, such as star 
forming galaxies and QSOs, seems to be required. 


\begin{center}
\begin{table*}
\caption[]{Comoving density of matter, $\rho_{\gamma \gamma}$ (in $M_{\odot} Mpc^{-3}$), exploded as $SN_{\gamma \gamma}$
and the number $N_{\gamma \gamma}$ (in $Mpc^{-3}$) of $SN_{\gamma \gamma}$ required to enrich the IGM at the critical 
metallicity of $Z = 10^{-4} Z_{\odot}$, for the three different IMFs. }
\begin{tabular}{l|l|l}
\noalign{\smallskip}
\hline
\hline
\noalign{\smallskip}
IMF &   $\rho_{\gamma \gamma}$     &   $N_{\gamma \gamma}$   \\
\noalign{\smallskip}
\hline
\noalign{\smallskip}
x=1.35  &   22922                 &    115     \\
x=0.95  &   22589                 &    113      \\
x=0.50  &   21923                 &    110  \\
\hline					
\hline					 
\end{tabular}				
\end{table*}
\end{center}


\begin{table*}
\begin{flushleft}
\centering
\caption[]{Average number of ionizing photons per star, total number of ionizing 
photons emitted by a generation of pop III stars per $Mpc^{3}$ and number of ionizing photons per baryon, 
for the three different IMFs.}
\begin{tabular}{l|l|l|l}
\noalign{\smallskip}
\hline
\hline
\noalign{\smallskip}
IMF &   $<N_{ion}>$     &    $N_{ion} (Mpc^{-3})$   & $N_{ion}/n_b$  \\
\noalign{\smallskip}
\hline
\noalign{\smallskip}
x=1.35  &  $2.28 \cdot 10^{64}$   &    $6.50 \cdot 10^{66}$  & 0.98     \\   
x=0.95  &  $2.80 \cdot 10^{64}$   &    $1.07 \cdot 10^{67}$  & 1.62   \\	 
x=0.5   &  $3.55 \cdot 10^{64}$   &    $1.07 \cdot 10^{67}$  & 1.62   \\		
\hline					
\hline					 
\end{tabular}				
\end{flushleft}
\end{table*}

\subsection{Predicted abundances of the intergalactic medium at high redshift}

Songaila (2001) observed the abundances of C and Si at redshift $z=5$ 
and found
$2 \times 10^{-4} C_{\odot}$ for C and $3.5 \times 10^{-4} 
Si_{\odot}$ for Si (where the solar abundances are by number and taken from
Anders \& Grevesse, 1989) $\footnote{The assumed solar abundances are : 
$3.3 \cdot 10^{-4}$ for carbon and $3.3 \cdot 10^{-5}$ for silicon}$. 
Although these abundances 
might be uncertain (at least by a factor of two)
due to uncertain ionization corrections, they 
are the only data available at redshift $z \ge 5$ at the moment.
The observed $(C/Si)_{IGM}$ ratio is hence $\sim 0.57
 \cdot (C/Si)_{\odot}$. 
We compute the yields per stellar generation of $^{12}C$ and $^{28}Si$,
as produced by $SN_{\gamma \gamma}$, and assumed, as valid in IRA, that the
abundance ratios are equal to the yield ratios.
In particular, the abundance ratio between two elements $X_{i}$ 
and $X_{j}$ is given by:
\begin{equation}
\frac{X_i}{X_j}=\frac{y_{i}}{y_{j}} 
\end{equation}
where $y_{i}$ and $y_{j}$ are the yields per stellar generation, 
as defined in equation \ref{yield}, 
referring to the amount of matter newly 
produced  and restored in the form of the elements  $i$ and $j$, 
respectively.
Therefore, we compare the predicted abundance ratios relative to 
the solar values with the  observed $(C/Si)_{IGM}=0.57 \cdot (C/Si)_{\odot}$. 
The values we have obtained
are between $0.013 \cdot (C/Si)_{\odot}$ and $0.011\cdot (C/Si)_{\odot}$, 
taking into account all 
the different yields and the different IMF indexes.  
If we steepen the IMF and adopt an index much larger than the Salpeter one 
we can obtain only a slightly larger 
value 
for the carbon to silicon ratio.   
We also tried to decrease the upper mass limit of pop III stars assuming
$M_{up}=200 M_{\odot}$ and found that the predicted $(C/Si)_{IGM}$, relative to the solar ratio, varies 
only slightly
(from 0.013 to 0.018 in the Salpeter case). In fact, the number of $SN_{\gamma\gamma}$ varies only by 10-15\%.
Hence, the only possible explanation of the abundance ratios observed by 
Songaila (2001) is that 
at redshift 5 there was already a substantial contribution to the carbon
abundances by intermediate mass stars of pop II.
Carbon, in fact, is thought to originate mainly in low and intermediate 
mass stars, as shown by Chiappini et al. (2003) in connection with the 
chemical 
evolution of the Milky Way.
In this case, in fact, the IRA is no more valid and detailed calculations 
of chemical evolution, taking into account stellar lifetimes, are necessary.
Another possibility is that Si is strongly depleted into dust grains 
and hence the real Si abundance is larger than quoted.
However, to reproduce the 
observed C/Si ratio in the IGM, we need to assume that  the real 
Si abundance should be much
larger than the observed one, which would imply an unlikely high dust 
content in the very metal poor early universe.

We should then conclude that at redshift $z=5$ the pop II stars have 
already ``erased'' the signatures of pop III objects.

\renewcommand{\baselinestretch}{1.0}
\begin{table*}
\centering
\caption{Abundance ratios generated by pop III with masses $12-270 M_{\odot}$ for the three different IMFs. 
}
\begin{tabular}{lcccccccccccc}
\\[-2.0ex] 
\hline
\hline
\\[-2.5ex]
\multicolumn{1}{l}{IMF}&\multicolumn{2}{c}{$[C/Fe]$}&\multicolumn{2}{c}{$[O/Fe]$}&\multicolumn{2}{c}{$[N/Si]$}&\multicolumn{2}{c}{$[Zn/Fe]$}\\
\multicolumn{1}{c}{}&\multicolumn{1}{c}{HW02}&\multicolumn{1}{c}{UN02}&\multicolumn{1}{c}{HW02}&\multicolumn{1}{c}{UN02}&\multicolumn{1}{c}{HW02}&\multicolumn{1}{c}{UN02}&\multicolumn{1}{c}{HW02}&\multicolumn{1}{c}{UN02}
\\
\hline
\hline
\\[-1.0ex]
x=1.35    &    -0.131 & -0.030  &  0.175 &  0.248 & -3.660 &  -1.829  &   
0.15   &  -0.40 \\             
x=0.95    &    -0.323 & -0.161  &  0.064 &  0.216 & -3.978 &  -2.241  &  
-0.30 &   -0.70  \\           
x=0.5     &    -0.469 & -0.295  & -0.025 &  0.171 & -4.341 &  -2.710  
&  -0.77  &   -1.07 \\             
\hline
\hline
\end{tabular}
\flushleft
\end{table*}


\renewcommand{\baselinestretch}{1.0}
\begin{table*}
\centering
\caption{Abundance ratios generated by pop III with masses $100-1000M_{\odot}$ for the three different IMFs. 
}
\begin{tabular}{lcccccccccccc}
\\[-2.0ex] 
\hline
\hline
\\[-2.5ex]
\multicolumn{1}{l}{IMF}&\multicolumn{2}{c}{$[C/Fe]$}&\multicolumn{2}{c}{$[O/Fe]$}&\multicolumn{2}{c}{$[N/Si]$}&\multicolumn{2}{c}{$[Zn/Fe]$}\\
\multicolumn{1}{c}{}&\multicolumn{1}{c}{HW02}&\multicolumn{1}{c}{UN02}&\multicolumn{1}{c}{HW02}&\multicolumn{1}{c}{UN02}&\multicolumn{1}{c}{HW02}&\multicolumn{1}{c}{UN02}&\multicolumn{1}{c}{HW02}&\multicolumn{1}{c}{UN02}
\\
\hline
\hline
\\[-1.0ex]
x=1.35    &    -0.44 & -0.35  &  0.04 &  0.19 & -5.700 &  -3.5  &   
-1.87   &  -1.91 \\             
x=0.95    &    -0.49 & -0.38  &  0.00 &  0.17 & -5.800 &  -3.5  &  
-1.86 &   -1.93  \\           
x=0.5     &    -0.54 & -0.42  &  0.04 &  0.16 & -5.800 &  -3.6  
&  -1.84  &   -1.92 \\             
\hline
\hline
\end{tabular}
\flushleft
\end{table*}

\subsection{Abundance ratios by pop III stars and DLAs}
DLAs are high redshift (many of them are at $z >3$) 
systems with observed metallicities larger than  $[Fe/H] \sim -2.5$. 
We compare the abundance ratios measured in 
low-metallicity DLAs with the 
ones predicted for pop III stars in order to infer whether DLAs carry the imprint of the chemical enrichment by the first 
generation of stars, as argued by some authors 
(e.g. HW02). 
In particular, we check if the DLA abundances can be explained by a 
pregalactic generation of pop III stars.

Here we consider only the DLAs with the lowest metallicities, i.e. the 
ones with $[Fe/H], [Si/H]  < -2.0$. 
In general, DLAs,  show solar proportions in their relative abundances already at 
these low metallicities.  
This is suggested 
by the study of individual DLAs at various metallicities 
(e.g. Dessauges-Zavadsky et al. 2004, Calura et al. 2003). 
Therefore, the implication is that it is unlikely that they 
could have retained the signature 
of pop III stars, with perhaps the exception of the elements produced 
in very massive stars. 
In fact, the interpretation is that DLAs are objects where 
the star formation proceeds slowly, thus showing the pollution from 
type Ia SNe already at low metallicities (see Calura et al. 2003).
Since dust depletion is a function of the metallicity (Vladilo 2002), in 
these metal-poor systems dust   
is likely  to play a negligible role in determining the abundance patterns,
and therefore we do not consider this effect here. 
Finally, since DLAs do not show metallicities lower than  
$[Fe/H] \sim -2.5$, i.e. much higher than the most plausible value for the 
critical metallicity for the transition from 
very massive ($m> 100 M_{\odot}$) to normal ($m< 100 M_{\odot}$) stars, 
we calculate the abundance ratios considering  
pop III stars with masses also in the range 
$12 M_{\odot} \le m \le 270 M_{\odot}$, 
(i.e. we include also the contribution by 
zero metallicity stars with mass $m < 100 M_{\odot}$.)
In table 4 we report the abundance ratios generated by pop III stars for 
various choices of the IMF and in the mass range 12-270$M_{\odot}$. 
\begin{figure*}
\centering
\vspace{0.001cm}
\epsfig{file=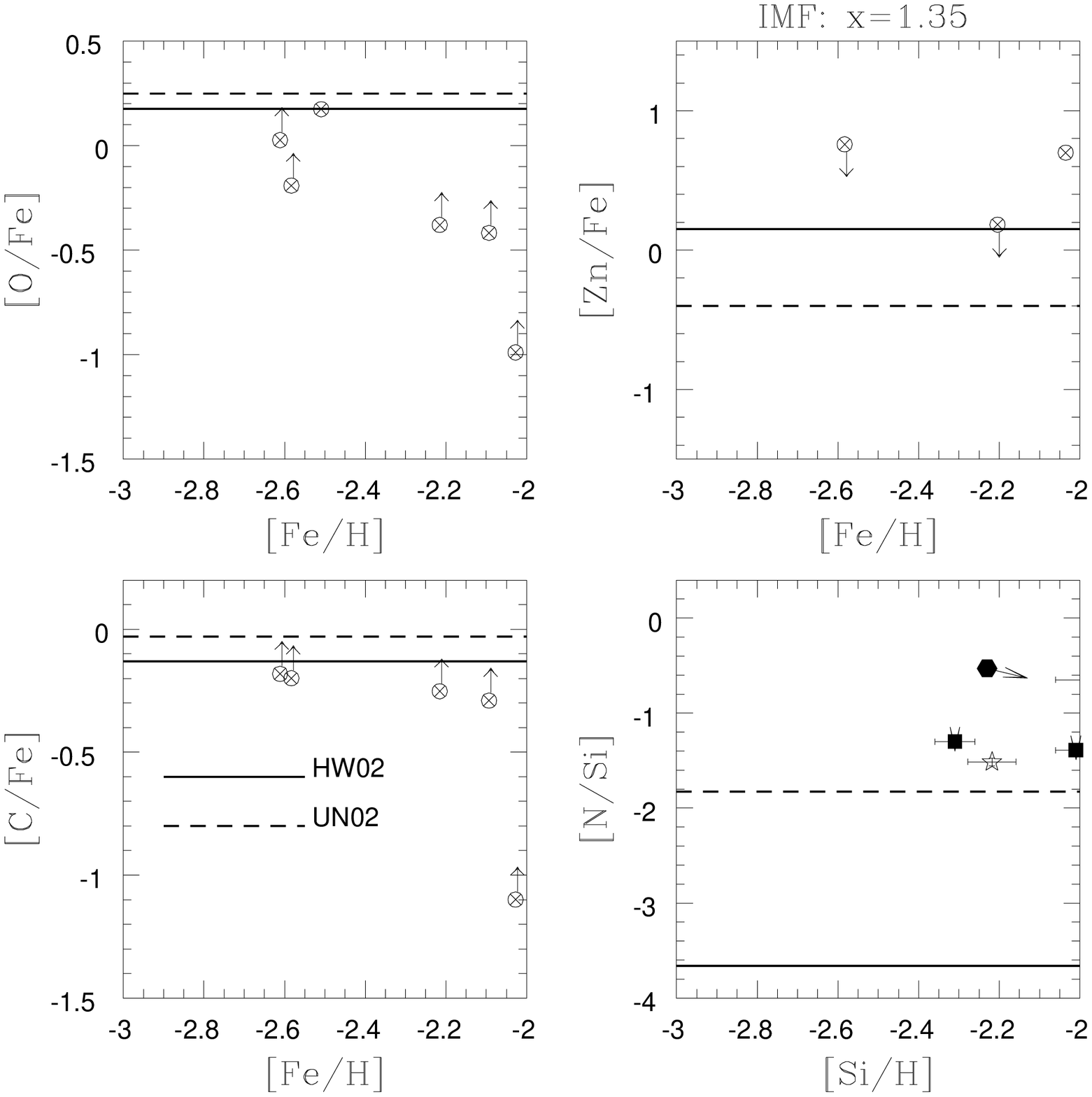,height=15cm,width=13cm}
\caption[]{Abundance ratios measured in low metallicity DLAs 
and predicted by adopting pop III stars with masses in the range 
$12 M_{\odot} < m < 270 M_{\odot}$ for the $x=1.35$ IMF. 
Points represent observations in DLAs, the dashed lines represent 
predictions obtained with UN02 yields for pop III stars, whereas the continuous lines represent the predictions obtained with HW02 yields.  
\emph{Crossed circles:} Prochaska \& Wolfe (2002); 
 \emph{five-points stars:} Prochaska et al. (2001); 
 \emph{hexagons:} Centuri\'on et al. (1998); 
 \emph{solid squares:} Pettini et al. (2002).  
All the DLAs have redshifts between $z=2$ and $z=3$.
}	
\end{figure*}

In Figure 2, we plot the abundance ratios measured in DLAs by various 
authors and the ones calculated for 
pop III stars in the case of a Salpeter IMF. All of the DLAs have low 
metallicity and redshifts between z=2 and z=3. 

In each panel, the continuous lines represent the predictions obtained 
with the yields of HW02, whereas the dashed lines indicate the predictions 
obtained with the yields of UN02. 
As one can see, in general the [O/Fe] ratio predicted by pop III yields is 
higher than the observed ones, although the data are all lower 
limits.
The predictions  for [C/Fe] and [Zn/Fe] with HW02 yields seem closer 
to the observed values than the UN02 yields. 
In the case of nitrogen, the pop III stars, irrespective of the different 
yields, seem to produce far less N than is observed. 
There is no way to reconcile the 
predicted values with observations even taking into account dust depletion. 
With a flatter IMF, as shown in table 4,
the disagreement would be even stronger. 
Therefore, as in the IGM at redshift 5, also in metal poor DLAs the 
observed patterns 
seem to have lost any trace of the early signature by pop III stars and to 
contain the products of pop II stars of intermediate mass. 
Low and intermediate mass stars ($0.8 \le m/m_{\odot} \le 8.0$), in fact, 
are believed to be the main N producers. In particular, at low metallicities 
N should be mostly produced as a primary element (e.g. by means of stellar rotation, as suggested by Meynet \& Maeder, 2002, or by dredge-up and hot bottom burning in AGB stars, as suggested by Renzini \& Voli, 1981).
Finally, in table 5 we show the same abundance ratios as in table 4 
but produced by a stellar generation made of only very massive stars.
As one can see, the agreement with DLA abundances is even worse.

We note that UN02 have already concluded that the abundance ratios 
produced by 
pop III stars do not agree with the abundance ratios measured in the 
halo stars of our Galaxy. 

Finally, we checked the effect of having zero-metal stars with 
masses only in the range 13-35$M_{\odot}$ (yields 
from Chieffi \& Limongi, 2004 
and UN02),
without $SN_{\gamma \gamma}$, and with a Salpeter IMF. In this case we 
find that
the predicted [C/Fe] and [N/Si] ratios are closer to those of the DLAs. 
However, to obtain this result we have to assume that the upper mass limit 
for the pregalactic stars was even lower than the present time one, 
which is a rather odd hypothesis.

\section{Discussion and conclusions}

In  this paper we have examined, by means of simple analytical 
calculations, the effects of a hypothetical stellar generation 
of very massive stars 
with no metals (population III stars), formed at very high redshift, on the 
chemical enrichment of the IGM at high redshift and on the chemical 
abundances of DLAs.

To do that, we have considered  stars with masses up to $10^{4} M_{\odot}$ and
taken into account the nucleosynthesis prescriptions for very massive 
pop III stars ($m> 100M_{\odot}$) from the most recent calculations.
We have explored different IMF prescriptions and in one case considered 
also stars in the range $12-100M_{\odot}$ of primordial chemical composition.

Our main conclusions are:
\begin{itemize}
\item It is impossible to reproduce the observed C/Si ratio in the IGM at 
redshift $z=5$ only with the contribution of very massive 
stars ($m >100 M_{\odot}$).
In particular, intermediate mass stars  (5-8$M_{\odot}$) 
of population II seem to be required to 
reproduce the carbon abundance. This 
is plausible given 
the short phase for the existence of pop III stars, which should form only 
up to a critical metallicity $Z_{cr}$ not higher than $10^{-4}Z_{\odot}$, 
a value reached rapidly in the IGM.

\item On the other hand, very few $SN_{\gamma \gamma}$ are required to enrich
the IGM at high redshift. This number depends only slightly on 
the assumed IMF.
In particular, to enrich a cubic megaparsec of the IGM at the level of the 
critical metallicity $Z_{cr}=10^{-4}Z_{\odot}$ the number of such SNe 
varies from 110 to 115 when the
index of the IMF varies between 1.35 and 0.5.
In the most favourable case,  with an IMF exponent $x=0.95-0.5$ 
we predict a number of ionizing 
photons per baryon of $\sim 1.6$, which is not enough to reionize the 
universe, although not negligible. It is worth noting that the number 
$SN_{\gamma \gamma}$ depends very weakly on the IMF exponent: this is due 
to the fact that we always normalize the IMF to unity (equation 2)
and the variation of the exponent $x$ is compensated by the variation of the 
normalization constant.

\item We have compared 
the predicted abundance ratios, [O/Fe], [C/Fe], [Zn/Fe] and [N/Si], produced 
by the nucleosynthesis in pop III stars with the same abundance ratios in low 
metallicity DLAs ([Fe/H]$< -2.0$) and concluded that the contribution 
of pop III stars alone cannot reproduce all of these ratios. In particular, 
it fails to reproduce the N abundance. 
This result suggests that the bulk of nitrogen in these objects
should arise from 
different stars, in
particular from pop II 
intermediate and low mass stars 
($0.8 \le M/M_{\odot} \le 8$).

\item We note that in a recent paper Matteucci \& Pipino (2004) 
explored the effects of pop III stars on the chemical evolution of spheroids.
They concluded that, if pop III stars formed only during a very short 
period of time (of the order of few million years), their effect on 
the stellar and gaseous abundances are negligible and one cannot 
assess their past existence or disprove it.
In the present paper we arrive at a similar conclusion
for the IGM at redshift z=5 and also for DLAs.   In fact, in these latter cases
we can only conclude that pop III stars are not enough to explain the observedabundances but we cannot exclude that $SN_{\gamma \gamma}$
ever existed.

\end{itemize}

\section*{Acknowledgments}
We wish to thank S. Cristiani, A. Ferrara and A. Pipino 
for helpful discussions and suggestions. We also like to thank an anonymous
referee who helped to improve noticeably the paper. 
This work has been supported by the italian MIUR
(Ministero Istruzione, Universita' e Ricerca) under COFIN03 
prot.2003028039.

\label{lastpage}


\begin{thebibliography}{99}
\bibitem[]{} Anders, E., Grevesse, N., 1989, Geochim. Cosmochim. Acta, 53, 197
\bibitem[]{} Bromm V., Larson R. B., 2004, ARA\&A, 42, 79
\bibitem[]{} Calura F., Matteucci F., Vladilo G., 2003, MNRAS, 340, 59
\bibitem[]{} Carr B. J., Bond J. R., Arnett W. D., 1984, ApJ, 277, 445
\bibitem[]{} Cen R., 2003a, ApJ, 591, L5
\bibitem[]{} Cen R., 2003b, ApJ, 591, 12
\bibitem[]{} Ciardi, B., Ferrara, A., White, S. D. M., 2003, MNRAS, 344, L7
\bibitem[]{} Centuri\'on, M., Bonifacio, P., Molaro, P., Vladilo, G., 1998, ApJ, 509, 620
\bibitem[]{} Chiappini C., Romano D., Matteucci F., 2003, MNRAS, 339, 63
\bibitem[]{} Chieffi A., Limongi M., 2004, ApJ, 608, 405 
\bibitem[]{} Dessauges-Zavadsky M., Calura F., Prochaska J. X., D'Odorico S., Matteucci, F., 2004, A\&A, 416, 79 
\bibitem[]{} El Eid M. F., Fricke K. J., Ober W. W., 1983, A\&A, 119, 54
\bibitem[]{} Ferrara, A., Salvaterra, X. 2004, in " Joint Evolution of Black Holes and Galaxies", M.Colpi, V.Gorini, F. Haardt, U. Moschella eds., in press, astro-ph/0406554
\bibitem[]{} Heger A., Fryer C. L., Woosley S. E., Langer N., Hartmann D. H, 2003, ApJ, 591, 288 
\bibitem[]{} Heger A., Woosley S. E., 2002, ApJ, 567, 532 (HW02)
\bibitem[]{} Matteucci, F., Pipino, A., 2004, MNRAS, in press
\bibitem []{} Meynet, G., Maeder, A., 2002, A\&A, 381, L25
\bibitem[]{} Ober W. W., El Eid M. F., Fricke K. J., 1983, A\&A, 119, 61 
\bibitem[]{} O'Meara, J. M., Tytler, D., Kirkman, D., Suzuki, N., Prochaska, J. X., Lubin, D., Wolfe, A. M., 2001, ApJ, 552, 718
\bibitem[]{} Pettini M., Ellison S. L., Bergeron J., Petitjean P., 2002, A\&A, 391, 21 
\bibitem[]{} Prochaska J. X., Wolfe A. M., Tyler D., Burles S., Cooke J., Gawiser E., Kirkman D., O'Meara J. M., Storrie-Lombardi L., 2001b, ApJS, 137, 21 
\bibitem[]{} Prochaska J. X., Wolfe A. M., 2002, ApJ, 566, 68
\bibitem[]{} Renzini, A., Voli, M. 1981, A\&A, 94, 175
\bibitem[]{} Schaerer, D. 2002, A\&A, 382, 28
\bibitem[]{} Schaye J., Aguirre A., Kim T.-S., Theuns T., Rauch M., Sargent W. L. W., 2003, ApJ, 596, 768
\bibitem[]{} Schneider R., Ferrara A., Natarajan P., Omukai K., 2002, ApJ, 571, 30
\bibitem[]{} Sokasian, A., Abel, T., Hernquist, L., 2003, MNRAS, 340, 473
\bibitem[]{} Songaila A., 2001, ApJ, 561, L153
\bibitem[]{} Spergel, D. N., et al., 2003, ApJS, 148, 175
\bibitem[]{} Tinsley, B.M. 1980, Fund. Cosmic Phys. 5, 287
\bibitem[]{} Umeda H., Nomoto K., 2002, ApJ, 565, 385, (UN02)
\bibitem[]{} Vladilo G., 2002, ApJ, 569, 295
\bibitem[]{} Woosley S. E., Weaver T. A., 1995, ApJS, 101, 181 
\end{thebibliography}
\end{document}